\documentclass[12pt]{article}
\usepackage{fullpage}

\begin{document}

\begin{flushright}
TIFR/TH/08-17
\end{flushright}
\bigskip

\begin{center}
\Large{\bf Scientific Realism and Classical Physics} \\
\bigskip\bigskip
\large{Virendra Singh} \\
\bigskip
INSA C.V. Raman Research Professor \\
Tata Institute of Fundamental Research \\
1, Homi Bhabha Road, Mumbai 400 005, India 
\end{center}
\bigskip

\begin{center}
\underbar{Abstract}
\end{center}
\bigskip\bigskip

We recount the successful long career of classical physics, from
Newton to Einstein, which was based on the philosophy of scientific
realism.  Special emphasis is given to the changing status and number
of ontological entitities and arguments for their necessity at any
time.  Newton, initially, began with (i) point particles, (ii) aether,
(iii) absolute space and (iv) absolute time.  The electromagnetic
theory of Maxwell and Faraday introduced `fields' as a new ontological
entity not reducible to earlier ones.  Their work also unified
electricity, magnetism and optics.  Repeated failure to observe the
motion of earth through aether led Einstein to modify the Newtonian
absolute space and time concepts to a fused Minkowski space-time and the
removal of aether from basic ontological entities in his special theory
of relativity.  Later Einstein in his attempts to give a local theory
of gravitation was led to further modify flat Minkowski space-time to
the curved Riemannian space time.  This reduced gravitational 
phenomenon to that of geometry of the space time.  Space-time, matter
and fields all became dynamical.  We also abstract some general
features of description of nature in classical physics and enquire
whether these could be features of any scientific description?

\newpage

\noindent 1. \underbar{Introduction}
\bigskip

Realism has been the dominant ontology among the practicing scientist
until at least the discovery of quantum mechanics in mid nineteen
twentities.  Indeed the origin of science among the ancient greeks
itself depended on it.  As Erwin Schr\"odinger, in his delightful
Shearman lectures given at University College at London in 1948 and
later printed as ``Nature and Greeks'', emphasized that Greeks based
their study of Nature on the following presuppositions:
\bigskip

\begin{enumerate}
\item[{(i)}] ``the hypothesis that the display of Nature can be
understood'', and 
\item[{(ii)}] ``the hypothesis of a real world around us''.
\end{enumerate}
\bigskip

The first of these leads one away from arbitrary mythological way of
thinking, and the second of these to the ``objectivation of the
world''.  These are essentially the creed of scientific-realism. 
\bigskip

The foundations of the classical physics, with which modern
development of physics begins, at the end of the medieval world, were
laid down by the great Isaac Newton (1642-1727).
\bigskip

\noindent 2. \underbar{Newtonian Mechanics}
\bigskip

Newtonian mechanical view of the physics takes the matter in the world
to consist of a number of ``absolutely hard indestructible
particles'', each endowed with a mass, moving on the stage of
unchanging three-dimensional space with time under the action of
mutual forces.  He also discovered the ``inverse square of the
distance'' law for the gravitational force between two bodies.  The
law was of an ``action at a distance form''.  This law of gravitation,
together with Newtons three laws of dynamics, gave a remarkably
accurate description of planetary motion. He thus laid the foundations
of his system of the world in his magnum opus ``Principia'' (1687).
\bigskip

Newton also postulated an elastic medium ``aether'' to pervade the
entire space and all the bodies contained theirin.  It's density was
taken to be greatest in the interplatenatry space and variable in the
various bodies.  It was also, on the analogy of water vapour in the
air, taken to contain various `aetherial spirits', suitable to produce
the phenomenon of electricity, magnetism and even gravitation. 
\bigskip

Unlike Hooke and others who took light to be waves the medium `aether',
Newton rejected the conception of light as waves.  His two main
arguments for the rejection were (i) the propagation of light rays in
straight lines, and (ii) the phenomenon of polarisation of light which
he had observed.  Another reason as to why Newton did not favour wave
nature for light was his theory of colours arising out of his
experiments on light refraction through glass prisms.  They proved to
him that different colours are already present in white light before
it is incident and these are not produced due to it's refraction.
Hooke had regarded colours to be produced due to the effect of medium
of wave propagation.  This not being the case Newton rejected the
Hooke's wave hypothesis for light as well.  The light for Newton thus
had a corpuscular nature.  The light and aether however affected each
other. 
\bigskip

When we observe the collision of two bodies on earth, we see a slowing
down of the motion due either to inelasticity of the bodies or
friction.  Nowadays we attribute this to a conversion of mechanical
energy to other forms of energy.  But, in Newton's time, such a
principle of conservation of energy had not yet been formulated.  So
for Newton, another reason for the `aether' was it's need to avoid 
the slowing down of the observed planetary motions over long durations
of time. 
\bigskip

Thus Newtonian mechanics, with which classical physics begins, and
which is one of it's magnificient achievements, subscribes to
realism.  It's ontology has (i) massive particles, (ii) space, (iii)
time and (iv) aether as the basic real entities. 
\bigskip

\noindent 3. \underbar{Wave theory of light}
\bigskip

The observation of the burning power of convex glasses at their focus,
however, suggested to Christiaan Huygens strongly that light must be
basically a wave motion.  The burning implies that light is associated
with some kind of motion.  In this experiment light beams moving in
different direction do not obstruct each other in any way but are
rather reinforcing in their effect which is possible with wave motion
in a medium but unlikely for projectiles moving in different
dimensions.  He therefore came out strongly in favour of wave theory
of light.  He also made great advances in the mathematical theory of
wave propagation.  He reported all these researches to the French
Academy in 1678.  Light was now regarded as waves in the universal
medium aether.
\bigskip

Further strong support for wave theory of light came in 1801 when
Thomas Young explained ``Neweton's rings'' on the wave theory, and
later using the same ideas the colours of thin films.  He clearly
formulated the general laws of interference of light waves and his
``double slit'' interference experiment is one of the celebrated
experiments in history of physics.  As interference phenomenon is not
possible to understand on the corpuscular theory of light, it provides
strong evidence for the wave theory.  The mathematical treatment of
diffraction of light, using wave theory was achieved by Auguste Jean
Frencl in his prize memoir submitted to French Academy in 1816.  He
did this by combining Huygen's work with Young's laws of interference.
\bigskip

The first tentative ideas that the polarisation of light is due to
``the light waves being transverse'' were given by Young in 1817 in
his attempt to explain Fresnel and Arago's experiments on interference
of polarised light.  Previously light was taken, in analogy to the
sound waves, as longitudinal.  Since only longitudinal waves are
possible in a fluid, it implied that aether must be an elastic solid.
Fresnal, around 1821, tried to deduce dynamical consequences
transverse waves for a solid aether.
\bigskip

After the work of Young, Fresnel, Arago and others the wave theory of
light found general acceptance.  The next great advance in the theory
of light came from an apparently totally unrelated area of studies on
electric and magnetic forces.

\bigskip\bigskip

\noindent 4. \underbar{The Electromagnetic field}
\bigskip

The earlier work on electrical and magnetic forces was modelled on
``the action at a distance'' Newtonian theory of gravitation.  A drastic
change of view point took place with Michael Faraday.  He discovered
the law of electromagnetic induction in 1831.  In trying to physically
understand the electromagnetic induction, Faraday began to think in
terms of ``lines of magnetic forces''.  These are the kind of curves
in which iron filing arrange themselves around a magnet.  He imagined
the density of these lines of forces at any point in space to
represent the strength of the magnetic intensity and their direction
as it's direction.  He thus imagined the whole space around the magnet
filled with a new entity ``magnetic field'' with a value and direction
at every point.  He had a similar conception of ``electric field''
among the space around any ``electrically charged object''.  We need
not now regard the electric force between two distant charged
particles as an ``action at a distance''.  Rather each particle is
surrounded by an electric field and it is this electric field to which
the other charged particles elsewhere respond.  We thus have a ``local
interaction'' theory of electrical, and similarly for magnetic forces.
\bigskip

Faraday did not express himself in mathematical language but as
Maxwell percieved ``his (Faraday) methods of conceiving the phenomenon
was also mathematical one, though not exhibited in the conventional
form of mathematical symbols''.  Together with a novel concept,
``displacement current'', a special flash of genius, Maxwell succeded
in distilling all the known laws of electricity and magnetism, into a
coherent mathematical structure, ``Maxwell's equations'' in 1864.
What emerged was the unification of the concept of electric and
magnetic into ``electromagnetic field''.
\bigskip

One can now similarly concieve gravitational forces also arising from
the local interactions via a gravitational field.  Thus was added
another class of ontological real entities ``fields'' to the classical
physics. 
\bigskip

An unexpected and remarkable fall out of the Maxwell's work was the
existence of waves of the electromagnetic field.  The velocity of
these waves, which was related to electric and magnetic polarisability
of the ``aether'', on being measured, was found to be numerically
equal to the known ``velocity of light'' in space.  Further they were
transverse in nature.  Maxwell then identified these electromagnetic
waves with light.  They were taken to be the waves in the same medium,
named ``luminiferous aether'', which was supportly responsible for
electromagnetic fields. 
\bigskip\bigskip

\noindent 5. \underbar{The existence of atoms; A challenge to realism}
\bigskip

John Dalton laid foundations of modern chemistry in 1808 through his
system of a finite number of chemical elements.  The indivisible
constituents of these elements were their respective atoms.  The
molecules were taken as composed of these atoms of chemical elements.
Amedeo Avogadro proposed in 1811, that equal volumes of any gas, under
same pressure and temperature conditions, contain the same number of
molecules.  This number is now called Avogadro Number for a mol. of
gas.  To be meaningful this law implies that molecules really exist.
In the nineteenth century most chemists used atoms and molecules as
heuristic devices to bring order into the description of chemical
reactions etc.  They did not necessarily believe in their existence. 
\bigskip

The development of kinetic theory of gases in the later half of
nineteenth century, where the gases were regarded as a system of
molecules in motion, took rapid strides.  Clausius in 1857 identified
heat as a form of molecular motion.  Maxwell and Boltzmann proposed
statistical mechanics methods to bring thermodynamical phenomenon
under the domain of Newtonian mechanics.  These efforts gave a big
boost to atoms being regarded as real entities.
\bigskip

The entire evidence for atoms was indirect as they were not seen
directly.  They gave rise to first serious challenge to realism in
classical physics.  The great physical chemist William Ostwald, as
well George Helm, regarded atoms to be just mathematical constructs.
This is quite analogous to situation about quarks in mid twentieth
century.  Ernst Mach, in view of his positivistic philosophy, also
doubted that atoms really exist.
\bigskip

This challenge to the reality of the atoms however was decisively
settled in favour of their existence through the theoretical
investigations on Brownian motion of small grains in the liquids carried by
Albert Einstein in 1905 and their subsequent experimental verification
by Jean Perrin in 1908-1913.  Perrin measured the Avogadro Number in
these investigation rather accurately.  If you can count their number
in a volume accurately, then the molecules must be real. 
\bigskip\bigskip

\noindent 6. \underbar{The motion of earth through aether}
\bigskip

\noindent 6.1 {\it Galilean Relativity}:
\medskip

Newton's three laws of motion are valid in a special set of reference
frames ie `inertial frames of reference'.  Newton's first law says
that a particle will keep moving in a straight line or remain at rest,
if it is not acted upon by any external force.  Clearly a particle
which is moving in a straight line in an earth laboratory will not
appear to be moving so if seen from another frame in which Sun, for
example, is at rest due to diurnal rotation and the annual revolution
around the Sun of the earth.  It is therefore obvious that both the frames,
geocentric and heliocentric ones, can not be inertial.  We have to
know, before using the Newton's laws, if the frame in which we are
applying them is inertial.
\bigskip

We first note a symmetry under transformation of space and time
coordinates, called ``Galilean relativity'' of the Law's of Newton.
If they are valid in a frame of reference S then they are also valid
in another frame of reference S$^\prime$ which is moving uniformly in a
straight line with respect to S.  Thus if S is an inertial frame then
so is the frame S$^\prime$.  This specifies the class of all equivalent
inertial frames of reference provided we can identify one of these
frames as inertial.  In practice this frame can be taken as the frame
in which the centre of mass of the solar system is at rest or has a
uniform linear motion.  Within accuracy required in the planetary
calculation it could well be one in which the system of fixed stars is a
rest or uniform linear motion.  Galilean relativity also specifies the
rules for transformating space and time, known as Galilean
transformations from one frame to another frame of reference. 
\bigskip

\noindent 6.2 {\it Maxwell's electromagnetic theory of light and
Galilean relativity}
\medskip

As long as Newton's laws of motion are the only fundamental laws of
nature, there is clearly no way to find the absolute velocity of any
system of reference with respect to some absolutely fixed point at
rest, in view of their invariance under Galilean relativity.  The
situation is drastically changed if we admit Maxwell's equations for
electromagnetic field, and light, to be fundamental as well.
\bigskip

Maxwell's equations are not invariant under Galilean transformations.
It is clearly seen by the fact that they predict the light velocity
$c$ to be a fixed number.  It can not be so in all inertial frames if
they are related to each other by Galilean transformation.  If the
light velocity is $\vec c$ in its direction of propagation in the
frame $S$ it would be given by $(\vec c + \vec v)$ in other frame
S$^\prime$ moving with respect to $S$ with a velocity $\vec v$.  The
velocity of light is thus a fixed constant only in the frame in which
luminiferous aether is at rest.
\bigskip

This clash of invariance of Newton's laws and noninvariance of
Maxwell's equations opens up the posibility to measure the velocity of
earth through the aether.  A large number of experiments were devised
for this purpose.  However all of them were unable to reveal any motion
of earth through the aether.  Most accurate and well known of these
experiments were those of Michelson and Morley in 1887.
\bigskip

\noindent 6.3 {\it Einstein's special theory of relativity}
\medskip

This conundrum of nondetection of the earth velocity through the
aether was attributed to possible dynamical effects such as
Fitzgerald and Lorentz length contraction hypothesis ie all bodies
shrink in length along their direction of motion by a certain factor
depending on their velocity to light-velocity ratio, but remain
unchanged in the other directions orthogonal to that of motion.
Einstein however completely changed our way of looking at the problem
through a revision of Newtonian concepts of absolute space and
absolute terms.  He reduced the problem to a change in kinematics. 
\bigskip

In his revision of the classical concepts of space and time,
Einstein's central point of departure was his analysis of the concept
of ``simultaneity of two events''.  Suppose we observe two events
taking place in a particular fixed frame of reference, say a railway
platform, then there is no difficulty in saying whether they are
simultaneous or not.  However if the same two events, which were
simultaneous when viewed from the railway platform, are viewed from
another frame moving with a velocity, such that of a railway train or
a linear track, they would not appear simultaneous.  This is because
the light signals from the two events, which are used to observe them,
will take different times to arrive at the observer in the moving
train in view of finite speed of light signals.  Since simultaneity is
not an invariant concept it follows that time can not be absolute.
\bigskip

As his guiding principle in finding the new concept of space and time
Einstein held on to the following two postulates:
\medskip

\noindent 1. \underbar{The principle of relativity}: All physical laws
have the same form in all inertial frames ie frames of reference which
move rectilinearly with a constant velocity with respect to each
other. 
\medskip

\noindent 2. The velocity of light is same in all inertial forms.
\bigskip

It was the crucial insight of Einstein to realise that while these
``two postulates of special relativity'', look irreconcilable, they are
indeed not so.  They appear so only if Newtonian concepts of space and
time are used.  Indeed if they are modified, so as to accord with the
relativity of simultaneity, they can be reconciled.  They then
support a new structure of fused space and time, called Minkowskian
space-time, instead of a separate space and a separate time
continuums.  It further follows that the two inertial frames are no
longer connected by Galilean transformations of space and time
coordinates and time is not an invariant.  They are now related by
Lorentz transformations.
\bigskip

A purely kinematic consequence of Lorentz tranformations is the
Fitzgerald-Lorentz length contraction, which was postulated earlier in
an ad-hoc way to explain the nondetection of earth's motion through
aether.  Another consequence is Einstein time dilation according to
which moving unstable particles such as pions live longer.
\bigskip

Maxwell's equations turn out to be invariant under Lorentz
transformations.  Newtonian laws of mechanics are however not so as
they are invariant under Galilean transformations.  Einstein therefore
proposed to modify them so as to make them also invariant under
Lorentz transformations.  Since, for velocities small compared to
velocity of light, the Lorentz transformations reduce to Galilean
transformation, the modified laws reduce to Newton's Laws in that
limit.  For higher velocities the expressions of the momentum and
energy of a particle to its velocity does change.  A far reaching
consequence is the Energy-mass equivalence.
\bigskip

It should be emphasised that as a result of special relativistic no
signals using physical particles, with a finite rest mass, can travel
faster than the velocity of light.
\bigskip\bigskip

\noindent 6.4 {\it The end of the ``aether''}
\medskip

As both relativistic-mechanics of Einstein and Maxwell's equations are
invariant under the same Lorentz transformation, the possibility of
measuring the velocity of earth through the aether is no longer there.
Not only that now that the velocity of light in \underbar{all}
inertial frames is the same there is no distinguished frame of
reference, which hitherto was referred to that as that of luminiferous
aether.  We can thus dispense with the concept of aether on that
score. 
\bigskip

Another reason on which aether was felt necessary in classical physics
was that it provided material substratum for the earliest
``phenomenological'' fields, such as mass-density and those corresponding
to matter-velocity, temperature.  All of these describe states of
matter inside massive bodies.  However to introduce these fields all
over space and time it was felt that some thing, named aether, is there
in the space even when massive matter is not present.  This was
especially so as ultimately all physical explanations were believed to
be mechanical in nature.  In fact originally Maxwell tried to reduce
the electromagnetic fields also to a mechanism of pulleys and gears
etc. but these attempts had to be given up as unsuccessful.  It
emerged that the concept of electromagnetic field necessitates that
the fields
are a new kind of ontological entity.  The crutch of aether for
imagining fields also gradually was not needed.  Electromagnetic field
could very well exist in space empty of matter.
\bigskip

As a result of the work on special theory of relativity the concept of
aether became superfluous and was dropped from physics.
\bigskip\bigskip

\noindent 7. \underbar{Gravitation and General Theory of Relativity}
\medskip

Further far reaching changes in our view of space-time, which is
Minkowskian in special theory of relativity, and gravitation were
still to come.
\bigskip

Newtonian theory of gravitation is an ``action at a distance'' theory
in which the gravitational effects propagate instantaneously.  This
does not conform to the tenets of special relativity.  It has to be
therefore modified.  Just as electromagnetic field is represented by
vector field, the Newtonian gravitation can be represented by a scalar
field.  It is easy to write a Lorentz-invariant scalar field equation
which would bring Newtonian theory in accord with special theory of
relativity.  This however did not satisfy Einstein.
\bigskip

It was known that the inertial mass of a body and it's gravitational
mass, ie the mass appearing in the force law of gravitation, are
exactly equal.  From this it follows that accelerations of a body due
to gravity is independent of it's nature.  Since inertial mass on a
body depends on its energy content, it is also independent of it's
energy.  Since these facts have no natural explanation in the special
theory of relativity.  Einstein felt that to have a proper
understanding of gravitation, one will have to generalise the special
theory of relativity.
\bigskip

Now special theory of relativity still has a set of special frames of
reference, called ``inertial frames'', which are equivalent for
describing nature.  In fact Ernst Mach has asked the question as to
why these frames are special.  There was no satisfactory answer to
this in special theory of relativity as there had been none in
newtonian mechanics.  In Nov. 1907, Einstein sitting at patent office
in Bern, while preparing a review article on relativity, had ``the
happiest thought of his life'' that a person, in free fall, does not
feel the force of gravity.  He converted this insight into his
``principle of equivalence'' in 1911: the phenomenon happening in an
uniformly accelerated laboratory are indistinguishable from those
happening in a uniform gravitational field.  Later it was generalised
to more general gravitational fields.  The final formulation of
Einstein's theory of gravitation, General theory of relativity, was
achieved in 1915.
\bigskip

In General theory all the reference frames, connected to each other by
differentiable coordinate transformations, are regarded as equally
good for description of nature, thus addressing the concern of Mach.
The laws of nature are thus form-invariant under such transformation
and not just under Lorentz transformation.  The space-time is no
longer Minkowskian but curved Riemannian.  The presence of matter
causes space-time to develop curvature.  The particles move in
geodesics in this curved space time.  The curvature effects of the
space-time are equivalent to the gravitational force effects.  The
metric tensor of the Riemannian space-time is identified with
gravitational field, which is now a symmetric second rank tensor and
not a scalar.  The principle of equivalence for general gravitational
fields gets now reformulated as follows: The Riemannian space-time is
locally transformable to a Minkowski one under coordinate
transformations.  The gravitational force is thus totally reduced to
the geometry of space-time.  With General relativity Einstein provided
a fitting capstone to the beautiful edifice of classical physics.
Space-time is no longer a passive stage on which matter plays it's
role, since matter and space-time react on each other, it is an active
participant. 
\bigskip

The general theory of relativity of Einstein is regarded as most
beautiful theory of physics.  The power of pure thought achieved so
much with so little experimental input.
\bigskip\bigskip

\noindent 8. \underbar{Some general features of World in the classical
physics}  
\medskip

Basing itself firmly on the rock of scientific-realism, classical
physics has had a glorious career.  The basic entitities are
point-particles, fields and space-time.  All of these obey dynamical
laws.  The world of classical physics is deterministic.  The role of
probability is only to model complex situations as in Brownian motion
and statistical mechanics so as to provide a useful simpler
description.  That determinism does not always lead to full
predictability for all time was a discovery of Poincar\'e in his study
of nonlinear chaotic systems. 
\bigskip

The theories of classical physics are theories in which all
interactions are local.  This is so for Maxwell-Faraday
electromagnetic interactions and Einstein's theory of general
relativity realised such a description for gravitation as well.
\bigskip

Another important and appealing feature of the classical physics is
it's unitary nature.  All physical systems, whether those which are
being measured, or those which are being used to measure, are
described by the same classical physics.  Within classical physics
this point was so obvious that it was not even pointed out.
\bigskip

Whether these features of classical description of nature are peculiar
to it or are they a deeper feature of any scientific description of
nature?  It may be tempting to answer the question raised with a
`yes', but a definitive answer, even about the validity of framework
of scientific realism, is not yet in.  As we know for atomic and
radiation problems one had to develop, during 1900-1925, a new
description, `Quantum Mechanics' as classical physics proved
inadequate for this task.  The answer to the question raised depends
on the problems of interpretation of quantum formalism, which is not
settled yet.  Depending on which interpretation we choose we get
different answers.
\bigskip\bigskip

\begin{center}
\underbar{Bibiliographical notes}
\end{center}
\bigskip

\begin{enumerate}
\item[{1.}] The two presuppositions about Greek Science are quoted
from Chapter 7 of \\ E. Schr\"odinger, \underbar{Nature and Greeks},
Cambridge Univ. Press, 1954. 
\item[{2.}] A very good source on the thought of Newtons and his
predecessors is \\  E.A. Burtt, \underbar{The Metaphysical Foundations
of Modern Science}, Humanities Press (1952), (sixth reprint 1992).
\item[{3.}] We also rely on the rather technical work, 
E. Whittaker: \underbar{A History of the theories of} \underbar{Aether and
electricity, Vol. 1 (The Classical Theories)},  \\ Harper, New York,
reprint 1960) \\ for a history of classical physics upto 1900.
\item[{4.}] For a history of atomism, See B. Pullman: \underbar{The
Atom in the history of human thought}, Oxford Univ. Press (1998).
\item[{5.}] A good scientific biography of Einstein, is A. Pais,
\underbar{Subtle is the Lord, ....; The Science} \underbar{and Life of Albert
Einstein}, Oxford Univ. Press, New York, 1973; \\ See also Einstein's
autobiographical notes in P.A. Schilpp (Editor), \underbar{Albert
Einstein,} \underbar{Philosophor-Scientist}, Harper, New York, 1949; \\ and
Einstein's ``Notes on the origin of the general theory of relativity''
in \\ A.P. French, \underbar{Einstein, A Centenary Volume}, Harvard
Univ. Press, Cambridge, 1879.
\item[{6.}] For a further treatment of the question raised in the last
section, see Virendra Singh, Bohm's realist interpretation of Quantum
Mechanics, TIFR preprint (2008).
\end{enumerate}

\end{document}